**Structure and Born effective charge determination for planar-zigzag β-poly(vinylidene fluoride) using density-functional theory**

*Nicholas J. Ramer\* and Kimberly A. Stiso*

*Department of Chemistry, Long Island University – C. W. Post Campus,*

*Brookville, New York 11548-1300*

**Abstract**

Two structures have been proposed in the literature for the β-phase of the ferroelectric polymer, poly(vinylidene fluoride) (β-PVDF); planar-zigzag and alternatively-deflected forms. Using density-functional theory, we have found the planar-zigzag structure is the preferred form and upon atomic relaxation, the alternatively-deflected structure attains a structure very similar to the planar-zigzag structure. In order to better understand the atomic origin of the ferroelectricity in β-PVDF, we have for the first time determined the dynamic Born effective charges ($Z^*$) for the planar-zigzag structure using a Berry-phase approach. When compared to their nominal ionic values, the $Z^*$ show anomalous differences. Using these effective charges, we describe the polarity of the bonds with β-PVDF and show the extent of atomic-motion-induced (or dynamic) charge transfer within this ferroelectric material. In addition, our effective charges are different to previously-determined Mulliken charges, due to the inherent differences between static and dynamic charges.

Keywords: Ferroelectric polymers, Density-functional theory, Effective charges

\* *Corresponding author*. Address: Department of Chemistry, Long Island University – C. W. Post Campus, 720 Northern Boulevard, Brookville, New York, 11548-1300; Tel.: 516-299-3034, fax: 516-299-3944.

*E-mail address*: nramer@liu.edu (N. J. Ramer).

# 1. Introduction

Poly(vinylidene fluoride) or PVDF has long been known to possess piezoelectric and pyroelectric behavior [1,2]. The ferroelectric nature of the polarization in PVDF was first reported by Furukawa *et al.*[3]. PVDF exhibits a much stronger piezoelectric response than most other known ferroelectric polymers. As compared to ceramic piezoelectric materials, PVDF has a relatively low dielectric constant and low elastic stiffness. Combining these two properties results in high voltage sensitivity and low acoustic impedance for this material. Also possessing desirable physical properties (flexibility, low density and ease of fabrication), PVDF and its related co-polymers have become very attractive for use in solid-state device applications.

PVDF has a monomeric repeat unit of ($-CH_2-CF_2-$). There are four known crystalline forms of PVDF: $\alpha$ (Form II), $\beta$ (Form I), $\gamma$ (Form III) and $\delta$ phases (Form IV or II$_p$). The phases differ in crystallographic space group, number of formula units per unit cell, packing of polymer chains and patterning of $-CH_2-$ and $-CF_2-$ conformations. The observed conformations are as follows: *trans* (*t*) (F and H atoms at 180º to polymer chain) and *gauche*$^\pm$ ($g^\pm$) (H and F atoms at $\pm$ 60º to polymer chain) [4]. The angles observed experimentally are slightly different.

The $\alpha$-phase shows a $tg^+tg^-$ conformation pattern with four formula units per unit cell and monoclinic space group $P2_1/c$ ($C_{2h}^5$) [5,6]. The chains in $\alpha$-PVDF are packed in an anti-parallel manner, yielding a non-polar structure. The $\beta$-phase possesses all-*trans* conformations with two formula units per unit cell and an orthorhombic space group and parallel chain packing (see below). The crystal structure of the $\gamma$-phase has been characterized by either an orthorhombic unit cell with space group $C2cm$ ($C_{2v}^{16}$) or monoclinic unit cell with space group $Cc$ ($C_s^4$) [7,8]. Regardless of unit cell geometry, a $tttg^+tttg^-$ conformation pattern has been observed with eight formula units per unit cell. The designation of the δ-phase as $II_p$ represents its identification as the polar form of the non-polar α-phase. It possesses parallel packing of polymer chains with all its dipoles oriented in the same direction.

Two structures for β-PVDF have been proposed in the literature. Galperin *et al.* suggested planar-zigzag conformations for all carbons in the polymer (so-called all-*trans*) [9,10]. The planar notation refers to all the carbon atoms lying in the *yz* plane. An orthorhombic unit cell was proposed by Lando *et al.* containing fully planar-zigzag conformations (see Figure 1a) [11]. In their planar-zigzag structure, the fluorine-fluorine distance is reported as 2.56 Å, which is equal to the *c* lattice constant of the reported orthorhombic unit cell. This distance is 0.14 Å less than twice the van der Waals radius of fluorine ($2 \times 1.35$ Å $= 2.70$ Å). This observation would seem to refute the assignment of a planar-zigzag structure to β-PVDF. Lando *et al.* proposed a crowding of the fluorine atoms along the chain axis and that atomic thermal motion will be restricted along this axis. They confirmed their structure by studying nuclear magnetic resonance spectra for the second moment of a β-PVDF film as a function of the angle of the magnetic field and the draw direction.

Subsequently, a different orthorhombic unit cell was proposed by Hasegawa *et al.* also containing fully planar- zigzag conformations [12]. As an alternative explanation to fluorine crowding, they also determined a structure for β-PVDF with alternatively-deflected zigzag molecular conformations (see Figure 1b). The deflections cause a deformation of the carbon backbone chain in an alternating pattern. Specifically, the carbon atoms bonded to the fluorine atoms alternate in their deflection (±*x*) from the *yz* plane (see Figure 2a). These –$CF_2$– deflections were originally proposed by Galperin *et al.* and account for the smaller chain axis length (2.56 Å) as compared to twice the covalent van der Waals radius of fluorine [9]. Due to the deflections, the fluorine-fluorine distance is increased to 2.60 Å. Justification for their alternatively-deflected model was given by a slight improvement (4.5%) in the fit of their proposed structure to the X-ray diffraction data and by calculating the minimum intermolecular potential energy as a function of the –$CF_2$– deflection angle, $\sigma$ [13,14].

The ferroelectricity in PVDF results from the presence of the highly electronegative F atoms, which yield highly polar bonds with C. Coupled with the much less polar bonds between C and H, a spontaneous polarization in β-PVDF is generated along the *b*- or *y*-axis. By symmetry, no component of the spontaneous polarization will be found in either the *x* or *z* direction.

Atomic charges can provide a more fundamental and better understanding of the underlying basis for the spontaneous polarization along with various other physical phenomena, which are central to many solid-state measurements, such as IR spectrum analysis and chemical shifts in X-ray photoemission spectroscopy. However, the ubiquitous usage of atomic charges has also led to many definitions and approaches to their computation. These approaches unfortunately are not equivalent [15]. The atomic charge definitions can be categorized in either static, based upon the partitioning of the charge density into contributions from specific atoms, or dynamic, defined by the change in polarization created by atomic displacement [16]. Based on these different definitions, static and dynamic charges cannot be considered equivalent, although differences in their magnitudes can be considered as an indicator of the extent of atomic charge transfer within the material.

A static atomic charge relies on the idea that the charge associated with an isolated atom is well-defined. Specifically, the delocalized electronic charge density is mapped onto localized point charges for each atom. This mapping can be done unambiguously if and only if a "boundary can be drawn between ions so as to pass through regions which the electron density is small compared with the reciprocal of volume inclosed [17]." In the case of strong covalent bonding, as is the case in organic-based polymers, these boundaries are not clearly defined due to the continuous charge density.

Dynamical Born effective charges ($Z^*$) can be computed from differences in the spontaneous polarization due to small atomic displacements away from a relaxed structure. $Z^*$ for atom $m$ is defined as

$$Z_m^* = \frac{\Delta P_s}{u_m} = \frac{P_s^{\text{disp}} - P_s^{\text{undisp}}}{u_m} \qquad (1)$$

where $u_m$ is the displacement of atom $m$, $P_s^{\text{disp}}$ is the spontaneous polarization for the displaced structure and $P_s^{\text{undisp}}$ is the spontaneous polarization for the undisplaced structure. Unlike static charges, Born effective charges are rigorous and are observed experimentally [18,19,20,21]. Determination of Born effective charges in ferroelectric materials has focused on inorganic perovskite materials, which contain metal-oxygen bonding [22,23,24,25,26]. The determination of $Z^*$ in covalently bonded organic ferroelectric polymers has not been reported.

Since there is limited experimental evidence of the existence of the alternatively-deflected structure, a theoretical determination of its relative stability as compared to the planar-zigzag structure is warranted. From minimization of computed atomic forces, we will be able to determine relaxed atomic positions for the planar-zigzag and alternatively-deflected structures. Finally, we compute the Born effective charges for our planar-zigzag structure to examine the atomic origin of the material's ferroelectric behavior and interatomic charge transfer.

## 2. Theoretical methodology

We have applied density-functional theory (DFT) within the generalized-gradient approximation (GGA) [27,28] and periodic boundary conditions. Optimized pseudopotential [29] were generated using the OPIUM code [30] for use with a plane-wave basis set cut-off of 50 Ry [31]. Each pseudopotential was constructed to have less than 1 mRy basis-set error per angular momentum. To determine the accuracy of these pseudopotentials, density-functional calculations within the GGA for various small molecules were completed. From these calculations, bond lengths and vibrational frequencies were determined and found to be within the expected error range for the GGA as compared to experiment [32,33]. The electronic minimizations were done using a conjugate-gradient method coupled with blocked-Davidson iterative diagonalization method with Pulay density mixing [34,35,36,37]

For the planar-zigzag β-PVDF unit cell, the lattice constants [$a = 8.58$ Å, $b = 4.91$ Å, and $c = 2.56$ Å] and the space group $Cm2m$ ($C_{2v}^{14}$) as determined by Hasegawa *et al.* were used [12]. The unit cell contains two formula units ($Z$=2), one at the origin and the other at (½ ,½,0). Each formula unit contains one monomer unit.

For the alternatively-deflected β-PVDF unit cell, the lattice constants [$a$ = 8.58 Å, $b$ = 4.91 Å, and $c$ = 5.12 Å] as determined by Hasegawa *et al.* and the space group *Cc2m* ($C_{2v}^{16}$) were used [12]. The doubled *c* lattice constant is necessary to incorporate the –CF$_2$– deflections (see Figure 1b). Our assignment of this space group agrees with the findings of Duan *et al.*[38]. The unit cell contains two formula units (*Z*=2), one at the origin and the other at (½ ,½,0). Each formula unit contains two monomer units.

For the atomic-relaxation density-functional calculations, Brillouin zone integrations were done using a 4 × 4 × 4 Monkhorst-Pack *k*-point mesh [39]. Convergence testing showed that the error due to *k*-point sampling at this mesh was less that 0.001 eV in the total energy per unit cell. Using the 4 × 4 × 4 mesh yields 8 *k*-points in the irreducible wedge of the Brillouin zone. The initial atomic positions for the planar-zigzag and alternatively-deflected β-PVDF structures were taken from Hasegawa *et al.* [12]. Atomic relaxations using a quasi-Newton quench method were completed by minimizing the computed Hellmann-Feynman forces on all atomic positions that are not constrained by symmetry [28,40,41,42,43,44]. Converged or relaxed atomic positions were achieved when the forces on all atoms were less than 0.01 eV/Å.

To determine the dynamic effective charges, an increased *k*-point mesh density must be employed parallel to the direction of polarization. For β-PVDF, a 4 × 8 × 4 *k*-point mesh was used. The Berry-phase method was employed to compute these values [45,46,47]. The computation of Born effective charges requires the determination of the polarization difference between a displaced and undisplaced structure [28]. It has been shown that the $P_s$ is linear in atomic displacement to a good approximation [48]. Atomic displacements from the relaxed structure were made in the direction of the polarization (along the *y*-axis) of 0.1% of the *b* lattice constant. Displacements in the other directions were not made due to space group symmetries. The electronic wave functions for these displaced structures were then determined using density-functional theory and the spontaneous polarization found for each structure. Using these polarizations and the polarization for the undisplaced structure, the dynamic effective charges were found via equation (1).

## 3. Results and discussion

### 3.1. Structure stability of planar-zigzag and alternatively-deflected β-PVDF

Upon atomic relaxation of both the planar-zigzag and alternatively-deflected structures, we find the planar-zigzag structure is 0.018 eV per formula unit lower in energy than the alternatively-deflected structure. This small energy difference would indicate that an alternatively-deflected structure is not stable at 0 K. We find several structural features of our relaxed alternatively-deflected structure that are similar to those in the planar-zigzag structure. First, we find that the –$CH_2$– groups have lost their *xy* rotation and have become symmetric around the chain axis (see Figure 2b). Furthermore, the –$CF_2$– deflections have been reduced. We find a reduction of σ from 7° in the experimental structure to approximately 3° in our final relaxed structure. This indicates that despite the slightly better fit to the experimental data of Hasegawa *et al*., the alternatively-deflected structure is not the preferred structure for β-PVDF at 0 K. It is however possible that the alternatively-deflected structure is accessible due to thermal motions at the temperature at which the X-ray was obtained (150 K). Based upon this finding, we will restrict our further discussion of results to the planar-zigzag structure [49].

The relaxed atomic positions for the planar-zigzag β-PVDF structure are given in Table 1. In the planar-zigzag structure, C1 and C2 are the carbons attached to the H and F atoms respectively. Experimentally-determined atomic positions are also included [11,12]. Several factors must be considered in the comparison of the experimental and theoretical values. First, the experimental values are determined at a finite temperature above 0 K. This means that some amount of thermal broadening will be present in the assignment of the atomic positions. In contrast, the theoretical positions are considered to be at 0 K. In addition, the positions of lighter atoms (such as H) are more difficult to ascertain experimentally. Even with these considerations, our relaxed atomic positions fall well within the range of experimental values.

As a further means of comparison, Table 2 reports bond lengths and bond angles for the theoretical and experimental structures. For the experimental structures, values are given based upon the reported atomic positions and those used to fit the X-ray patterns and intensities (in parentheses). Since H atom positions were not given in the Lando *et al.* X-ray structure [11], C1–H bond length and H–C1–H bond angle are not given although fitting values are provided.

We find excellent agreement between our bond lengths and those of the later X-ray study [12]. However, in comparing with the earlier experimental study, we find a large difference in the C2–F bond length. The use of an ideal tetrahedral angle (109.5º) in the fitting of their X-ray data may have contributed to the differences in their atomic positions and bond lengths as compared to the present study and the later experiment.

*3.2. Spontaneous polarization and dynamic Born effective charges*

In order to assess dynamic charge transfer (or charge transfer due to atomic motion) within β-PVDF, Born effective charges must be determined. Using the theoretically determined atomic positions in Table 1 and the electronic wave functions obtained from density-functional calculations, the $P_s$ for the relaxed (or undisplaced) structure of planar-zigzag β-PVDF was found to be 0.181 C/m² per unit cell volume. Our value for the polarization is in excellent agreement with the value found by a previous DFT study using a similar Berry-phase approach (0.178 C/m² per unit cell volume) [50].

The $Z^*$ for planar-zigzag β-PVDF were found to be –0.20 for C1, +1.45 for C2, –0.75 for F and +0.14 for H. The $Z^*$ satisfy the acoustic sum rule $\sum_m Z^*_m = 0.00$ to 0.04 for the unit cell, indicating well-converged calculations. It must be recognized that $Z^*$ reflect the covalency of the bonding environment of each atom with respect to some reference or nominal ionic value. In the case of C, the reference value is unclear, due to the covalent C1–C2 bonding. For purposes of this discussion, we will assume the nominal ionic values for C1 and C2 are –2 and +2 respectively. The values for F and H will be –1 and +1, respectively. Based on these nominal values, we therefore have found anomalous $Z^*$ in β-PVDF, a feature found in other ferroelectric materials. The term anomalous is used to refer to $Z^*$ that differ from their nominal ionic values.

There are several observations that can be made from the effective charge results. The origin of the inequivalency of the $Z^*$ for the carbon atoms stems from their different chemical environments.

First, we find that the $Z^*$ for C2 and F are very close to their reference ionic charges. This is caused by the electron-withdrawing behavior of the F atoms. This results in a highly ionic bond. Furthermore, the large difference between these $Z^*$ indicates that there is a large amount of charge transfer between these atoms when the C2–F bond length is varied. The origin of this charge transfer lies in the accompanying redistribution of tightly-held $2p$ electrons on the F atoms when either atom is displaced.

Due to the strong covalency of the C1–H bonds we find $Z^*$ for these atoms are very different from their reference values. This results from the atoms' similar electronegativities and the high amount of covalent character in their bond. The small difference between these $Z^*$ indicates that there is a small amount of charge transfer between these atoms when the C1–H bond length is varied. The lack of lone pairs of electrons on the H atom (as compared to the F atoms) contributes to this effect.

As a means of gauging the amount of charge transfer that results from atomic displacement in β-PVDF, we can examine differences between our dynamic effective charges and static Mulliken charges. Static charges do not capture the effect of charge transfer due to atomic displacements. A previous determination of these static charges used 1,1,1,3,3-pentafluorobutane, a representative molecule for the β-PVDF structure [51]. This study found Mulliken charges for C1 to be –0.406, C2 to be +0.781, F to be –0.398 and H to be +0.171 [52]. In the case of the polar C2–F bond, we find $Z^*$ approximately half their corresponding Mulliken values indicating a high amount of charge transfer present. The Mulliken charges for C1 and H are closer to (although larger than) their Born values. It is important to note that the chain-terminating groups in 1,1,1,3,3-pentafluorobutane (–$CF_3$ and –$CH_3$) will have an effect on the finite chain's dipole moment. This may have also contributed to the differences between the Born effective and Mulliken charges [53].

A final implication should be considered when discussing difference between static charges (e.g. Mulliken, Bader or potential-derived) and dynamic charges. Born effective charges rigorously yield the correct spontaneous polarization for a material. This fact is crucial in molecular dynamics studies in which dipole fluctuations are used to compute dielectric response properties such as dielectric constants or effects due to an externally applied field. For example, Born effective charges for liquid water were computed from within a Car-Parrinello molecular dynamics simulation using a variant of the Berry-phase method (so-called "one-point Berry phase") [54]. The resulting infrared spectrum showed good agreement with experiment and some features previously unseen in theoretical spectra were found due to the incorporation of the effective dipole moments.

Incorporation of the calculation of Born effective charge determination within molecular dynamics calculations is computationally intensive since the calculation of the effective charges requires an increased *k*-point mesh. However, the use of recently formulated maximally localized Wannier functions (in the case of insulators, occupied orbitals) in molecular dynamics simulations could facilitate expedited calculations of the $Z^*$ [55]. Alternatively, a simple model could be developed for the dependency of effective charges on localized atomic motions, such as bond length or angle. This parameterized model could then be used to determine the Born effective charges as a function of molecular structure.

## 4. Conclusions

We have examined two proposed structures for the β-phase of the ferroelectric polymer, poly(vinylidene fluoride); planar-zigzag and alternatively-deflected using density-functional theory.  We have found the alternatively-deflected structure is not stable and upon relaxation, adopts a structure close to planar-zigzag.  We have determined the dynamic Born effective charges for the planar-zigzag structure of β-PVDF using a Berry-phase approach.  The charges are anomalously different than their nominal ionic values, similar to what is found in inorganic ferroelectric materials, although smaller in magnitude.  We find high ionic character in the C–F bonds and high covalent character in the C–H bonds.  Finally, our effective charges vary greatly from previous determined Mulliken charges, indicating a high amount of dynamic charge transfer present in this material, especially in the –$CF_2$– groups.

## Acknowledgements

The authors would like to thank Andrew M. Rappe, Ilya Grinberg and Valentino R. Cooper for helpful discussions.  This work was supported by a grant from the Research Committee of the C. W. Post Campus of Long Island University.  Funding for computational support was provided by Long Island University.

**Figures**

Fig. 1. Molecular structures for β-poly(vinylidene fluoride). (a) planar-zigzag and (b) alternatively-deflected zigzag structures. See text for description of structures.

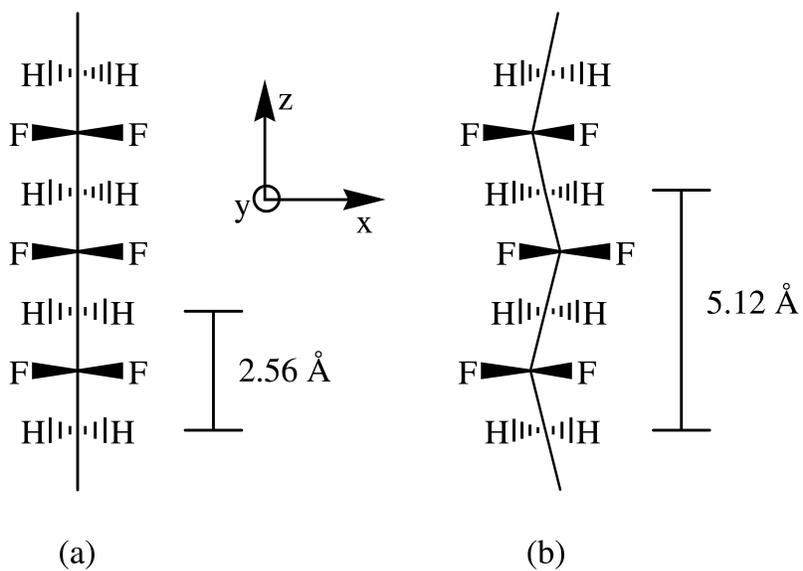

(a)  (b)

Fig. 2. Chain axis view of β-poly(vinylidene fluoride). (a) Alternatively-deflected zigzag structure with –CH$_2$– *xy* rotations and σ = 7° according to reference 12, (b) alternatively-deflected zigzag structure with no –CH$_2$– *xy* rotations and σ ≈ 3° found from atomic relaxation in the present study. σ is the deflection angle of carbon backbone.

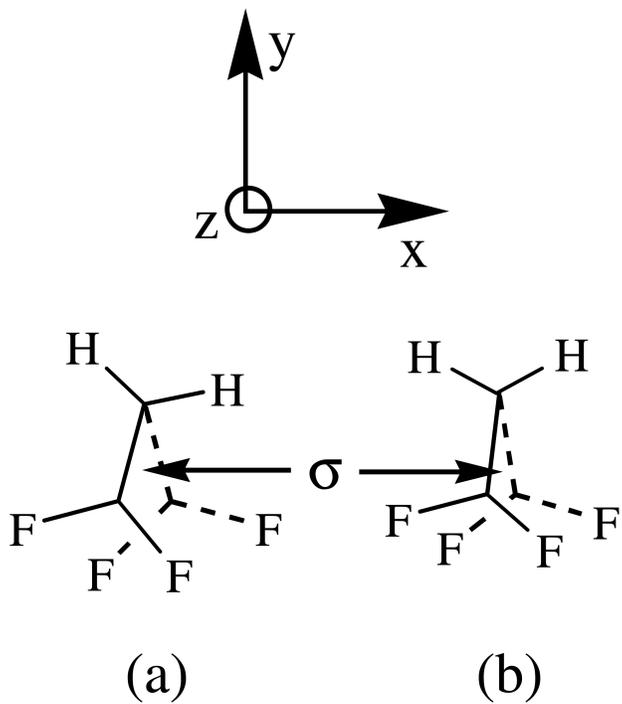

**Tables**

Table 1. Theoretical and experimental atomic positions for the planar-zigzag structure of β-poly(vinylidene fluoride). Positions are given as fractions of lattice constants.

|      | Theory[a] | | | Experiment[b] | | | Experiment[c] | | |
| --- | --- | --- | --- | --- | --- | --- | --- | --- | --- |
| Atom | x/a | y/b | z/c | x/a | y/b | z/c | x/a | y/b | z/c |
| C1 | 0.000 | 0.000 | 0.000 | 0.000 | 0.000 | 0.000 | 0.000 | 0.000 | 0.000 |
| C2 | 0.000 | 0.170 | 0.500 | 0.000 | 0.176 | 0.500 | 0.000 | 0.174 | 0.500 |
| F  | 0.129 | 0.343 | 0.500 | 0.130 | 0.334 | 0.500 | 0.126 | 0.355 | 0.500 |
| H  | 0.103 | -0.132 | 0.000 | n/a | n/a | n/a | 0.105 | -0.124 | 0.000 |

[a] $a = 8.58$ Å, $b = 4.91$ Å, and $c = 2.56$ Å.

[b] From reference 11 with $a = 8.47$ Å, $b = 4.90$ Å, and $c = 2.56$ Å.

[c] From reference 12 with $a = 8.58$ Å, $b = 4.91$ Å, and $c = 2.56$ Å.

Table 2. Theoretical and experimental bond lengths and bond angles for planar-zigzag structure of β-poly(vinylidene fluoride) computed from reported atomic positions. Values in parentheses indicate quantities used in fitting X-ray data to determine experimental atomic positions (See Table 1). Bond lengths are given in Angstroms and bond angles in degrees.

|  | Theory | Experiment[a] | Experiment[b] |
| --- | --- | --- | --- |
| Bond Lengths | | | |
| C1–H | 1.10 | (1.094) | 1.09(1.09) |
| C2–F | 1.39 | 1.35(1.344) | 1.40(1.34) |
| C1–C2 | 1.53 | 1.54(1.541) | 1.54(1.54) |
| Bond Angles | | | |
| H–C1–H | 107.5 | (109.5) | 111.9(112) |
| F–C2–F | 105.3 | 109.8(109.5) | 101.2(108) |
| C1–C2–C1 | 113.6 | 112.1(112.4) | 112.6(112.5) |

[a] From reference 11.

[b] From reference 12.

to the 50 Ry structure. In addition, the spontaneous polarization for the 60 Ry structure was identical to that of the 50 Ry structure.